\documentclass{llncs}
\usepackage{xcolor} 
\usepackage[normalem]{ulem} 
\usepackage{colortbl} 
\usepackage[T1]{fontenc}
\usepackage{float}
\usepackage{graphicx}
\usepackage{adjustbox}
\usepackage{hyperref}
\usepackage{multirow}
\usepackage{multicol}
\usepackage{bbding}
\usepackage{tabularx}
\usepackage{placeins}

\begin{document}
\title{Dual channel CW nnU-Net for 3D PET-CT Lesion Segmentation in 2024 autoPET III Challenge}
%
%
\author{Ching-Wei Wang\Envelope \and
Ting-Sheng Su \and
Keng-Wei Liu}
\authorrunning{F. Author et al.}
%
\institute{Graduate Institute of Biomedical Engineering, National Taiwan University of Science and Technology, Taipei, Taiwan\\
\email{cweiwang@mail.ntust.edu.tw}}
\maketitle              
\begin{abstract}

PET/CT is extensively used in imaging malignant tumors because it highlights areas of increased glucose metabolism, indicative of cancerous activity. Accurate 3D lesion segmentation in PET/CT imaging is essential for effective oncological diagnostics and treatment planning. In this study, we developed an advanced 3D residual U-Net model for the Automated Lesion Segmentation in Whole-Body PET/CT - Multitracer Multicenter Generalization (autoPET III) Challenge, which will be held jointly with 2024 Medical Image Computing and Computer Assisted Intervention (MICCAI) conference at Marrakesh, Morocco. Proposed model incorporates a novel sample attention boosting technique to enhance segmentation performance by adjusting the contribution of challenging cases during training, improving generalization across FDG and PSMA tracers. The proposed model outperformed the challenge baseline model in the preliminary test set on the Grand Challenge platform, and our team is currently ranking in the 2nd place among 497 participants worldwide from 53 countries (accessed date: 2024/9/4), with Dice score of 0.8700, False Negative Volume of 19.3969 and False Positive Volume of 1.0857. GitHub repository is available at \href{https://github.com/cwwang1979/CW-nnU-Net-for-PET-CT}{https://github.com/cwwang1979/CW-nnU-Net-for-PET-CT}

\keywords{Deep Learning,  whole-body PET/CT Imaging, Multi-tracer, 3D lesion segmentation, attention boosting.}
\end{abstract}
\section{Introduction}
Oncological diagnostics have been significantly enhanced by Positron Emission Tomography / Computed Tomography (PET/CT) imaging, which provides essential insights necessary for accurate diagnosis and effective treatment planning\cite{li2018use,lang2020impact}. Fluorodeoxyglucose (FDG) PET/CT is extensively used in imaging malignant tumors because it highlights areas of increased glucose metabolism, indicative of cancerous activity \cite{hirata2021quantitative,parihar2023fdg}. Besides FDG, prostate-specific membrane antigen (PSMA) is another crucial PET tracer used for the diagnosis and staging of prostate cancer, owing to its high expression on most prostate cancer cells \cite{farolfi2021current,combes2022psma}. A major barrier to clinical adoption of quantitative PET/CT analysis is lesion segmentation, a necessary step for detailed and individualized tumor characterization and therapeutic decision-making. Manual lesion segmentation is a labor-intensive and time-consuming process, making automated PET/CT lesion segmentation essential for enhancing efficiency, accuracy, and reproducibility.

Recent studies have demonstrated the potential of deep learning methods in the challenging task of automated PET/CT lesion segmentation, showing effective tumor segmentation across various research efforts. One approach derives attention maps from PET data to focus a convolutional neural network (CNN) on areas of the CT image with a higher likelihood of lung tumors \cite{fu2021multimodal}. Another method combines feature maps of different resolutions to create spatially varying fusion maps, enhancing lesion information for liver tumor segmentation \cite{xue2021multi}.  Furthermore, promoting the segmentation of head and neck oropharyngeal primary tumors in FDG-PET/CT images, the HECKTOR challenge has shown that a 3D U-Net based approach significantly exceeded human inter-observer agreement levels, highlighting the effectiveness of automated segmentation methods \cite{oreiller2022head}.

The MICCAI 2024 Automated Lesion Segmentation in Whole-Body PET/CT - Multitracer Multicenter Generalization Challenge (autoPET III) \cite{ingrisch_2024_10990932}, building upon the previous two challenges, i.e. 2022 autoPET and 2023 autoPET II, aims to meet the critical need for developing models capable of generalizing across various tracers and centers. The main objective is to accurately segment FDG- and PSMA-avid tumor lesions in whole-body PET/CT scans. The 2024 challenge provides FDG-PET/CT data from University Hospital Tübingen (UKT) and PSMA-PET/CT data from University Hospital LMU Munich (LMU), necessitating algorithms that minimize false-positive segmentation in anatomical structures with naturally high uptake. The complexity is heightened in a multitracer environment, as physiological uptake differs between tracers; for instance, FDG shows significant uptake in the brain, kidneys, and heart, whereas PSMA is more pronounced in the liver, kidneys, spleen, and submandibular glands.

In this study, we developed a 3D residual U-Net model with an enhanced architecture to increase depth and complexity, modifying both the encoder and decoder structures to improve feature extraction capabilities. Our training strategy incorporates a sample re-weighting technique, prioritizing more challenging samples to optimize model performance. Through iterative adjustments and retraining, this approach significantly improved segmentation accuracy and robustness.

\section{Method}
\subsection{Data}
The autoPET III challenge provides 1,611 studies in training dataset includes 1,014 FDG studies from 900 patients and 597 PSMA studies from 378 patients (see Table \ref{tab:materials}). The FDG group comprises 501 patients with histologically confirmed malignant melanoma, lymphoma, or lung cancer, along with 513 negative control patients. The PSMA group consists of PET/CT images of 378 male individuals with prostate carcinoma, with 537 images showing PSMA-avid tumor lesions and 60 images without. The FDG dataset from UKT includes 570 male patients and 444 female patients. In contrast, the PSMA dataset from LMU consists of 378 male patients. Imaging conditions differ between the datasets; the FDG Tübingen dataset was obtained using a single Siemens Biograph mCT scanner, whereas the PSMA Munich dataset was collected using three different scanners: Siemens Biograph 64-4R TruePoint, Siemens Biograph mCT Flow 20, and GE Discovery 690 \cite{fdgpet2022dataset,psmapet2024dataset}.

In the challenge preliminary evaluation, the test set is obtained from 5 studies in the released training dataset, includes 3 FDG studies and 2 PSMA studies. the final evaluation will be conducted using a hidden test set of 200 studies, comprising 50 FDG scans from LMU, 50 FDG scans from UKT, 50 PSMA scans from LMU, and 50 PSMA scans from UKT (see Table \ref{tab:materials}). 
Half of the test data will be drawn from the same sources and distributions as the training data, while the remaining half will be cross-sourced from the other center, ensuring a robust assessment of the model's generalization capabilities. The FDG PET/CT training and testing images from UKT were annotated by a radiologist with a decade of experience in hybrid imaging, while those from LMU were annotated by a radiologist with eight years of experience. The PSMA PET/CT datasets from both centers were initially annotated by a single reader and subsequently reviewed by a radiologist with five years of experience in hybrid imaging.

\begin{table}
\caption{Materials: autoPET III Challenge Dataset.}\label{tab:materials}
\begin{adjustbox}{width=\textwidth}
\centering
\begin{tabular}{|c|c|c|c|}
\hline
Source Sites                                                                                              & Training (Public)                   & Validation (Public)              & Testing (Private) \\ \hline
\multirow{2}{*}{LMU Hospital} & \multirow{2}{*}{1014 studies (FDG)} & \multirow{2}{*}{3 studies (FDG)}  & 50 studies (FDG)  \\ \cline{4-4} &    &  & 50 studies (PSMA) \\ \hline
\multirow{2}{*}{UKT Hospital}   & \multirow{2}{*}{597 studies (PSMA)} & \multirow{2}{*}{2 studies (PSMA)} & 50 studies (FDG)  \\ \cline{4-4}  &  &  & 50 studies (PSMA) \\ \hline
\end{tabular}
\end{adjustbox}
\end{table}

In total, 1611 studies are released and made available for challenge participants, and we categorized the FDG and PSMA PET/CT images from both institutions into two categories: images with tumor annotations (positive diagnoses) and those without tumor annotations (negative studies). For model training and internal evaluation, only the annotated images were utilized, amounting to a total of 1,038 images, namely Dataset P (see Table \ref{tab:trainingposneg}).

\begin{table}
\caption{Categorization of Training Studies: Positive and Negative}\label{tab:trainingposneg}
\centering
\begin{adjustbox}{width=\textwidth,totalheight={1cm},keepaspectratio}
\begin{tabular}{|c|c|c|}
\hline
All Training studies         & Positive (Dataset P)    & Negative (Non-utilized)   \\ \hline
1611 studies & 1038 studies & 573 studies \\ \hline
\end{tabular}
\end{adjustbox}
\end{table}

\subsection{Data Pre-processing}

For data normalization, 
we firstly analyzed the intensity values from the foreground classes (excluding the background and ignore classes) from all training cases by computing the mean, standard deviation, and the 0.5 and 99.5 percentiles of these values. Intensity values were then clipped to the 0.5 and 99.5 percentiles, followed by subtraction of the mean and division by the standard deviation \cite{isensee2021nnu}. This data normalization was applied to each training case separately for individual input channels, namely CT and PET channels.

\subsection{Data Augmentation}

In the proposed framework, two data augmentation strategies were explored and tested initially, including 
Type1 and Type2 Augmentation.

\paragraph{Type1 Augmentation}: The Type1 Augmentation includes basic transformations such as random rotation, cropping, scaling, Gaussian blur, Gaussian noise, contrast and brightness adjustments, down-sampling, and gamma correction. These augmentations are designed to enhance model robustness while maintaining the integrity of the original data distribution.

\paragraph{Type2 Augmentation}: The Type2 Augmentation extends the Type1 strategy with more complex augmentations, including enhanced [scaling, contrast and brightness], sharpening, localized gamma transformations and random rectangle occlusion simulations. These aim to expose the model to a wider range of possible data variations.

\paragraph{Performance Comparison}: By comparing the performance of Model2 and Model3, which are adopting Model1 as pre-trained weights, it is found that the Type1 Augmentation strategy  outperforms Type2 Augmentation (see Table \ref{tab:DataAugcomparision}). While Type2 Augmentation provides a broader range of augmentations, it can introduce variations that deviate too far from the actual data, potentially leading to poor performance. In contrast, the Type1 strategy better preserves the original data characteristics, resulting in improved model performance. As a result, in the following models, we adopted the Type1 Augmentation strategy instead of Type2 Augmentation.


\begin{table}
\centering
\caption{Performance Comparison on Type1 and Type2 Augmentation}\label{tab:DataAugcomparision}
\begin{adjustbox}{width=\textwidth}
\begin{tabular}{|c|c|c|ccc|}
\hline
\multirow{2}{*}{Model} & \multirow{2}{*}{\begin{tabular}[c]{@{}c@{}}Pre-trained\\ weights\end{tabular}} & \multirow{2}{*}{\begin{tabular}[c]{@{}c@{}}Data \\ Augmentation\end{tabular}} & \multicolumn{3}{c|}{Dataset P  (1038 files)}                                                         \\ \cline{4-6} 
                       &                                                                                &                                                                               & \multicolumn{1}{c|}{Dice score} & \multicolumn{1}{c|}{False Positive Volume} & False Negative Volume \\ \hline
Model1                 & -                                                                              & Type1                                                                         & \multicolumn{1}{c|}{0.8074}     & \multicolumn{1}{c|}{4.5071}                & 6.3643                \\ \hline
Model2                 & Model1                                                                         & Type2                                                                         & \multicolumn{1}{c|}{0.8048}     & \multicolumn{1}{c|}{4.8231}                & 7.0752                \\ \hline
Model3                 & Model1                                                                         & Type1                                                                         & \multicolumn{1}{c|}{0.8577}     & \multicolumn{1}{c|}{3.4134}                & 5.1799                \\ \hline
\end{tabular}
\end{adjustbox}
\end{table}

\subsection{Model Architecture}
Proposed model is built on a 3D residual U-Net architecture, manually adjusted to include a 7-stage encoder with convolutions arranged as (1,3,4,6,6,6,6). This enhances the model’s depth and complexity in the deeper stages. The decoder consists of 6 stages with one convolution per stage (1,1,1,1,1,1), designed to balance computational load while maintaining high performance. This setup deviates from the standard nnU-Net architecture, which typically includes a (1,3,4,6,6,6) encoder and a (1,1,1,1,1) decoder. Our modifications introduce an additional stage in both the encoder and decoder to create a deeper architecture.

The model processes 2-channel input data derived from CT and PET images. It utilizes a patch size of 128x256x256 and increases the maximum number of features from the standard limit of 320 to 384. This enhancement enables the model to capture more detailed features at deeper layers.

\subsection{Training Strategy}
Each model was trained with an initial learning rate of 0.0002. Additionally, we canceled the mirroring step in the data augmentation process, as the organs of the human body are not symmetrical. In our apporach, we developed a sample attention boosting model, which emphasizing difficult samples during training. First, we randomly divided all positive studies in the training set, Dataset P (1038 samples), into Dataset A (934 samples) and Dataset B (104 samples) to evaluate model performance. 


\begin{table}
\centering
\caption{Trainnig Dataset Division}\label{tab:datasetsplit}
\begin{adjustbox}{width=\textwidth}
\begin{tabular}{|l|cccc|}
\hline
\multirow{2}{*}{} & \multicolumn{4}{c|}{Dataset P (1038 samples)}                                                \\ \cline{2-5} 
                  & \multicolumn{2}{c|}{Dataset A (934 samples)}                               & \multicolumn{2}{c|}{Dataset B (104 samples)}          \\ \hline
Subset            & \multicolumn{1}{c|}{Dataset A*}       & \multicolumn{1}{c|}{Dataset A**} & \multicolumn{1}{c|}{Dataset B*}       & Dataset B** \\ \hline
Attribute         & \multicolumn{1}{c|}{lower Dice score} & \multicolumn{1}{c|}{remaining}   & \multicolumn{1}{c|}{lower Dice score} & remaining   \\ \hline
Number of samples   & \multicolumn{1}{c|}{50}               & \multicolumn{1}{c|}{884}         & \multicolumn{1}{c|}{54}               & 50          \\ \hline
\end{tabular}
\end{adjustbox}
\end{table}

We used Dataset A as the training set to trained Model1 for 1500 epochs, resulting in a Dice score of 0.8074. By analyzing the Dice score of each sample from Model1 results, we further divided Dataset A and Dataset B as follows: Dataset A* with 50 samples and Dataset B* with 54 samples, both representing lower Dice samples, and Dataset A** with 884 samples and Dataset B** with 50 samples, both representing remaining samples (see Table \ref{tab:datasetsplit}). To improve the model's accuracy, in Model2, we identified the 50 samples with lower Dice scores in Model1 and boosting the attention weights of these samples in the training set (see Table \ref{tab:modeltrainingsamples}). Additionally, we added the 54 samples with lower Dice scores from Dataset B** into the training set. The aim was to increase the model attention onto the samples with poor prediction performance in the training set, with the goal of improving performance. We subsequently employed Model1 as the pre-trained weights and re-trained it for 1000 epochs, while also modifying the data augmentation strategy from the Type1 to Type2 Augmentaion. However, the performance of Model2 did not improve significantly, with Dice score of 0.8048. 

\begin{table}
\caption{Detail of training samples for each model.}\label{tab:modeltrainingsamples}
\begin{adjustbox}{width=\textwidth}
\centering
\begin{tabular}{|c|cccc|c|c|c|}
\hline
\multirow{2}{*}{Model} & \multicolumn{4}{c|}{number of extra samples}                                                                         & \multirow{2}{*}{\begin{tabular}[c]{@{}c@{}}samples from\\ Dataset A\end{tabular}} & \multirow{2}{*}{\begin{tabular}[c]{@{}c@{}}samples from\\ Dataset B\end{tabular}} & \multirow{2}{*}{total samples} \\ \cline{2-5}
                       & \multicolumn{1}{c|}{Dataset A*} & \multicolumn{1}{c|}{Dataset A**} & \multicolumn{1}{c|}{Dataset B*} & Dataset B** &                                                                                 &                                                                                 &                              \\ \hline
Model1                 & \multicolumn{1}{c|}{0}          & \multicolumn{1}{c|}{0}           & \multicolumn{1}{c|}{0}          & 0           & 934                                                                             & 0                                                                               & 934                          \\ \hline
Model2                 & \multicolumn{1}{c|}{50}         & \multicolumn{1}{c|}{0}           & \multicolumn{1}{c|}{0}          & 0           & 934                                                                             & 54                                                                              & 1038                         \\ \hline
Model3                 & \multicolumn{1}{c|}{43}         & \multicolumn{1}{c|}{0}           & \multicolumn{1}{c|}{0}          & 0           & 934                                                                             & 54                                                                              & 1031                         \\ \hline
Model4                 & \multicolumn{1}{c|}{43}         & \multicolumn{1}{c|}{17}          & \multicolumn{1}{c|}{5}          & 0           & 934                                                                             & 104                                                                             & 1103                         \\ \hline
Model5                 & \multicolumn{1}{c|}{61}         & \multicolumn{1}{c|}{51}          & \multicolumn{1}{c|}{9}          & 0           & 934                                                                             & 104                                                                             & 1159                         \\ \hline
\end{tabular}
\end{adjustbox}
\end{table}


In Model3, we did not boost sample attention weights of the samples with Dice score of 0. Instead, we boosted sample attention weights of 43 samples from Dataset A* and added 54 samples from Dataset B*. The Model1 was again used as a pre-trained weights, and the model was re-trained for 1000 epochs, achieving Dice score of 0.8577. In Model4, the Model4 training set continued from the Model3 training set. By evaluating the Dice scores of Model3 using Dataset A** and Dataset B*, we identified samples with Dice scores below 0.7: 17 samples from Dataset A** and 5 samples from Dataset B*. We further boosted sample attention weights of the identified samples in the training set. Additionally, we performed a continuous training based on Model3's weights for 500 epochs, resulting in Dice score of 0.8715. 


The Model5 training set continued from the Model4 training set. By evaluating the Dice scores of Model4 using Dataset A*, Dataset A** and Dataset B, we identified samples with Dice scores below 0.75: 18 samples from Dataset A*, 34 samples from Dataset A** and 4 samples from Dataset B. We had subsequently boosted sample attention weights of the identified samples in the training set. Furthermore, we performed a continuous training based on Model4's weights for 500 epochs, resulting in Dice score of 0.8781.

\subsection{Post-processing}
No post-processing is performed for high computational efficiency.

\section{Results}
We evaluated the performance of our segmentation model using three evaluation metrics: the foreground Dice score, False Positive volume (FPvol) and False Negative volume (FNvol). The foreground Dice score evaluates the accuracy of the segmentation by quantifying the overlap between the predicted segmented lesions and the actual lesions by manual annotations. The FPvol metric measures over-segmentation by determining the volume of incorrectly identified lesion regions that do not overlap with the reference annotations. In contrast, the FNvol metric gauges under-segmentation by calculating the volume of actual lesion regions that are not captured by the model’s segmentation, where these regions do not overlap with the predicted mask. These metrics provide a thorough evaluation of the model's segmentation performance.

In Table \ref{tab:Overallresult}, the performance metrics of Model1 on the Dataset P were as follows: Dice score of 0.8074, FNvol of 4.5071 and FPvol of 6.3643. Through series of optimizations from Model 1, 3, 4 to 5, including a sample attention boosting model, Type1 data augmentation and continuous training, we achieved significant improvements in Model5: Dice score of 0.8781, FNvol of 2.4250 and FPvol of 4.4920.

\renewcommand{\arraystretch}{1.2}
\begin{table}[H]
\centering
\small
\caption{Overall Model Performance}
\label{tab:Overallresult}
\begin{tabular}{|c|c|c|c|c|c|c|c|c|c|}
\hline
 & \multicolumn{3}{c|}{Dataset A (934 samples)} & \multicolumn{3}{c|}{Dataset B (104 samples)} & \multicolumn{3}{c|}{Dataset P (1038 samples)} \\ \hline
 & Dice & FPvol & FNvol & Dice & FPvol & FNvol & Dice & FPvol & FNvol \\ \hline
Model1 & 0.8229 & 4.6176 & 6.3638 & 0.6686 & 3.5148 & 6.3680       
       & 0.8074 & 4.5071 & 6.3643 \\ \hline
Model2 & 0.8095 & 5.0585 & 7.2096 & 0.7621 & 2.7093 & 5.8679       
       & 0.8048 & 4.8231 & 7.0752 \\ \hline
Model3 & 0.8607 & 3.6268 & 5.3192 & 0.8313 & 1.4965 & 3.9281       
       & 0.8577 & 3.4134 & 5.1799 \\ \hline
Model4 & 0.8718 & 3.2122 & 5.0088 & 0.8690 & 1.1432 & 3.3856 & 0.8715 & 3.0049 & 4.8462 \\ \hline
Model5 & \boldmath{\textbf{0.8779}} & \boldmath{\textbf{2.6001}} & \boldmath{\textbf{4.6241}} & \boldmath{\textbf{0.8801}} & \boldmath{\textbf{0.8521}} & \boldmath{\textbf{3.3054}} & \boldmath{\textbf{0.8781}} & \boldmath{\textbf{2.4250}} & \boldmath{\textbf{4.4920}} \\ \hline
\end{tabular}
\end{table}

In Tables \ref{tab:DatasetAresult} and \ref{tab:DatasetBresult}, the Dice scores of Model1 with evaluation using Dataset A*, i.e. samples with the lower Dice score, and using Dataset B* were 0.3261 and 0.4967, respectively. By boosting the attention weights of the samples with the lower Dice scores by the preliminary model, the model was better trained by putting extra focus on poorly segmented cases, thereby improving overall performance. Consequently, the Dice scores of Model5 evaluated on samples with lower Dice scores by the previous Model improved significantly, i.e. from 0.3261 to 0.6485 for Dataset A* and from 0.4967 to 0.8471 for Dataset B*.

\begin{table}[H]
\centering
\caption{Quantitative Result of samples with lower Dice (A*) and remaining samples (A**)}
\label{tab:DatasetAresult}
\begin{tabularx}{\textwidth}{|c|>{\centering\arraybackslash}X|>{\centering\arraybackslash}X|>{\centering\arraybackslash}X|>{\centering\arraybackslash}X|>{\centering\arraybackslash}X|>{\centering\arraybackslash}X|}
\hline
\multicolumn{1}{|c|}{}&\multicolumn{6}{c|}{Dataset A}\\ \hline
\multicolumn{1}{|c|}{} & \multicolumn{3}{c|}{Dataset A* (lower Dice 50 samples)} & \multicolumn{3}{c|}{Dataset A** (remaining 884 samples)} \\ \hline
& Dice & FPvol    & FNvol  & Dice  & FPvol & FNvol    
\\ \hline
Model1  & 0.3261  & 31.3092 & 49.7958 & 0.8510 & 3.1079 & 3.9073    \\ \hline
Model2  & 0.4155  & 29.4534 & 50.2960 & 0.8318 & 3.6797 & 4.7726    \\ \hline
Model3  & 0.5721  & 24.3287 & 39.1188 & 0.8770 & 2.4559 & 3.4075    \\ \hline
Model4  & 0.6319  & 22.9165 & 36.9530 & 0.8853 & 2.0960 & 3.2020    \\ \hline
Model5  & \textbf{0.6485}   & \textbf{17.8155} & \textbf{30.5988}           & \textbf{0.8908}   & \textbf{1.7395}  & \textbf{3.1550}    \\ \hline
\end{tabularx}
\end{table}

\begin{table}[H]
\centering
\caption{Quantitative Result of samples with lower Dice (B*) and remaining samples (B**)}
\label{tab:DatasetBresult}
\begin{tabularx}{\textwidth}{|c|>{\centering\arraybackslash}X|>{\centering\arraybackslash}X|>{\centering\arraybackslash}X|>{\centering\arraybackslash}X|>{\centering\arraybackslash}X|>{\centering\arraybackslash}X|}
\hline
\multicolumn{1}{|c|}{}&\multicolumn{6}{c|}{Dataset B}\\ \hline
\multicolumn{1}{|c|}{} & \multicolumn{3}{c|}{Dataset B* (lower Dice 54 samples)} & \multicolumn{3}{c|}{Dataset B** (remaining 50 samples)} \\ \hline
& Dice  & FPvol    & FNvol     & Dice    & FPvol     & FNvol        \\ \hline
Model1  & 0.4967 & 3.7272 & 9.9517 & 0.8542 & 3.2854 & 2.4976     \\ \hline
Model2  & 0.6939 & 2.7421 & 8.9123 & 0.8357 & 2.6739 & 2.5800       \\ \hline
Model3  & 1.0561 & 5.2380 & 12.5645& 0.8792 & 3.5535 & 2.5135       \\ \hline
Model4  & 1.0111 & 3.8111 & 4.4965 & 0.8965 & 1.2860 &\textbf{1.7538}                
\\ \hline
Model5  & \textbf{0.8471} & \textbf{0.8636} & \textbf{4.7301}      
        & \textbf{0.9157} & \textbf{0.8398} & 1.7667    
\\ \hline
\end{tabularx}
\end{table}

In Table \ref{tab:preliminarysubmit}, the \textbf{datacentric\_baseline} refers to the official baseline model provided by the autoPET III challenge organizers, as listed in the preliminary test set leader board on the Grand Challenge platform. Our initial Model1, trained on the partial positive training dataset A, obtained Dice score of 0.8402, FNvol of 26.2889 and FPvol of 1.4888. 
Through the proposed attention boosting method and continuous training strategy, a significant improvement is observed in Model5, achieving Dice score of 0.8700, FNvol of 19.3969 and FPvol of 1.0857. Our team is currently ranking in the 2nd place among 497 participants worldwide from 53 countries (accessed date: 2024/9/4).


\begin{table}[H]
\centering
\caption{Preliminary Test Set Result in the Leader board  (accessed date: 2024/9/4)} 
\label{tab:preliminarysubmit}
\begin{tabular}{|c|c|c|c|c|}
\hline
Rank & Model                & Dice score & False Negative Volume & False Positive Volume \\ \hline
3rd  & Model5               & \textbf{0.8700}     & \textbf{19.3966}               & \textbf{1.0857}                \\ \hline
4th  & Model4               & 0.8632     & 20.4623               & 1.0334                \\ \hline
7th  & Model3               & 0.8462     & 21.1228               & 1.6206                \\ \hline
9th  & Model1               & 0.8402     & 26.2889               & 1.4888                \\ \hline
15th & datacentric\_baseline & 0.7990     & 23.8380               & 2.0503                \\ \hline
\end{tabular}
\end{table}

\begin{figure}[H]
    \centering
    \includegraphics[width=0.8\textwidth]{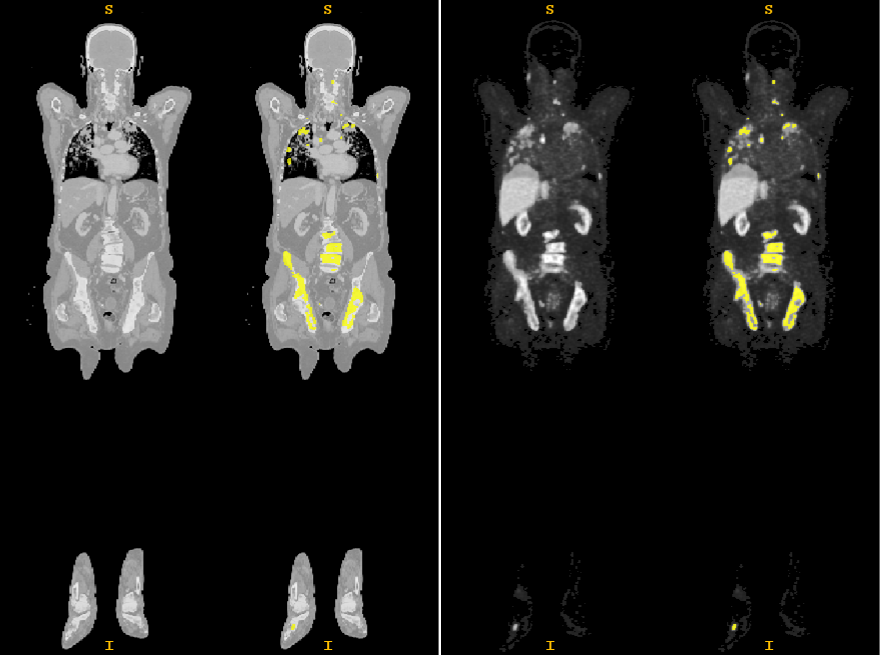}
    \caption{Qualitative whole-body lesion segmentation results using CT and PET Images CT (left) and PET (right) with yellow segmentation masks}
    \label{fig:CTPET}
\end{figure}

\begin{figure}[t]
    \centering
    \includegraphics[width=0.8\textwidth]{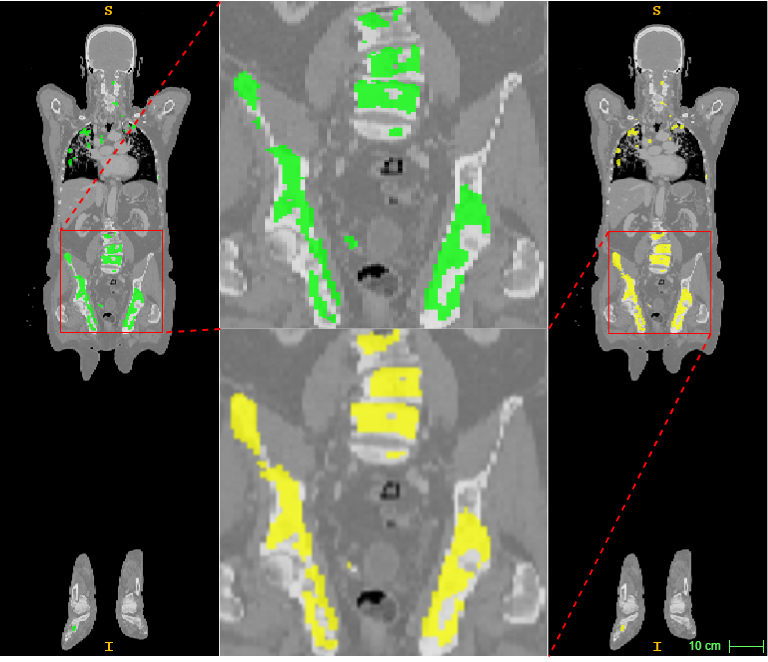}
    \caption{Whole-body lesion segmentation results on CT images. The left side displays the manually annotated ground truth lesions (in green), while the right side shows the predicted segmentation results from the model (in yellow). The middle section provides a zoomed-in view of the lesion areas.}
    \label{fig:CT_gt_pred}
\end{figure}

\begin{figure}[H]
    \centering
    \includegraphics[width=0.8\textwidth]{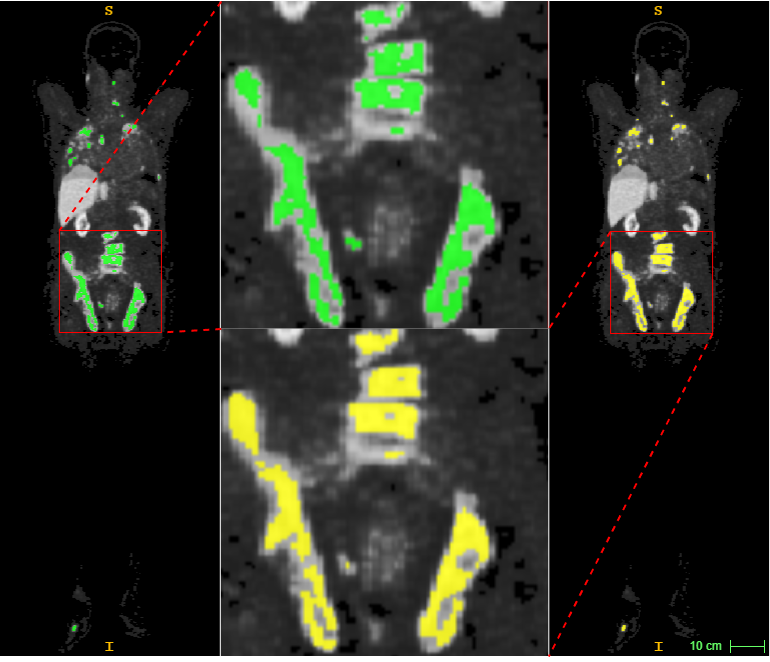}
    \caption{Whole-body lesion segmentation results on PET images. The left side displays the manually annotated ground truth lesions (in green), while the right side shows the predicted segmentation results from the model (in yellow). The middle section provides a zoomed-in view of the lesion areas.}
    \label{fig:PET_gt_pred}
\end{figure}

Figure \ref{fig:CTPET},\ref{fig:CT_gt_pred},\ref{fig:PET_gt_pred}  present the whole-body lesion segmentation results using CT and PET images generated by the proposed method, accompanied by a qualitative analysis comparing the predicted segmentations with the ground truth annotations. Through qualitative analysis, it can be observed that the predicted results demonstrate a high level of consistency with the ground truth in terms of lesion segmentation, indicating good model performance. Particularly in lesion boundary and region identification, the predicted segmentation outcomes closely match the manual annotation results.

\section{Discussion}

To address the challenge of PET/CT data from multiple tracers and centers, we proposed a dual channel CW 
nnU-Net with a 
sample attention boosting model, achieving low FPvol, low FNvol and high Dice score. 
This technique emphasizes difficult samples during training, adjusting their contribution to improve the model's performance. The results from the preliminary test set on the Grand Challenge platform demonstrated the effectiveness of the proposed training strategy (see Table \ref{tab:preliminarysubmit}). 

\section{Acknowledgement}
This study is supported by the National Science and Technology Council of Taiwan (NSTC 113-2222-E-011-003-MY3).


%
%
%
%

\bibliographystyle{splncs04}
\bibliography{main}

\begin{thebibliography}{10}
\providecommand{\url}[1]{\texttt{#1}}
\providecommand{\urlprefix}{URL }
\providecommand{\doi}[1]{https://doi.org/#1}

\bibitem{li2018use}
Li, R., Ravizzini, G.C., Gorin, M.A., Maurer, T., Eiber, M., Cooperberg, M.R.,
  Alemozzaffar, M., Tollefson, M.K., Delacroix, S.E., Chapin, B.F.: The use of
  pet/ct in prostate cancer. Prostate cancer and prostatic diseases
  \textbf{21}(1),  4--21 (2018)

\bibitem{lang2020impact}
Lang, D., Wahl, G., Poier, N., Graf, S., Kiesl, D., Lamprecht, B., Gabriel, M.:
  Impact of pet/ct for assessing response to immunotherapy—a clinical
  perspective. Journal of Clinical Medicine  \textbf{9}(11), ~3483 (2020)

\bibitem{hirata2021quantitative}
Hirata, K., Tamaki, N.: Quantitative fdg pet assessment for oncology therapy.
  Cancers  \textbf{13}(4), ~869 (2021)

\bibitem{parihar2023fdg}
Parihar, A.S., Dehdashti, F., Wahl, R.L.: Fdg pet/ct--based response assessment
  in malignancies. Radiographics  \textbf{43}(4),  e220122 (2023)

\bibitem{farolfi2021current}
Farolfi, A., Calderoni, L., Mattana, F., Mei, R., Telo, S., Fanti, S.,
  Castellucci, P.: Current and emerging clinical applications of psma pet
  diagnostic imaging for prostate cancer. Journal of Nuclear Medicine
  \textbf{62}(5),  596--604 (2021)

\bibitem{combes2022psma}
Combes, A.D., Palma, C.A., Calopedos, R., Wen, L., Woo, H., Fulham, M., Leslie,
  S.: Psma pet-ct in the diagnosis and staging of prostate cancer. Diagnostics
  \textbf{12}(11), ~2594 (2022)

\bibitem{fu2021multimodal}
Fu, X., Bi, L., Kumar, A., Fulham, M., Kim, J.: Multimodal spatial attention
  module for targeting multimodal pet-ct lung tumor segmentation. IEEE Journal
  of Biomedical and Health Informatics  \textbf{25}(9),  3507--3516 (2021)

\bibitem{xue2021multi}
Xue, Z., Li, P., Zhang, L., Lu, X., Zhu, G., Shen, P., Shah, S.A.A., Bennamoun,
  M.: Multi-modal co-learning for liver lesion segmentation on pet-ct images.
  IEEE Transactions on Medical Imaging  \textbf{40}(12),  3531--3542 (2021)

\bibitem{oreiller2022head}
Oreiller, V., Andrearczyk, V., Jreige, M., Boughdad, S., Elhalawani, H.,
  Castelli, J., Valli{\`e}res, M., Zhu, S., Xie, J., Peng, Y., et~al.: Head and
  neck tumor segmentation in pet/ct: the hecktor challenge. Medical image
  analysis  \textbf{77},  102336 (2022)

\bibitem{ingrisch_2024_10990932}
Ingrisch, M., Dexl, J., Jeblick, K., Cyran, C., Gatidis, S., Kuestner, T.:
  {Automated Lesion Segmentation in Whole-Body PET/CT - Multitracer Multicenter
  generalization} (Apr 2024). \doi{10.5281/zenodo.10990932}

\bibitem{fdgpet2022dataset}
Gatidis, S., Kuestner, T.: A whole-body fdg-pet/ct dataset with manually
  annotated tumor lesions (fdg-pet-ct-lesions) [dataset] (2022).
  \doi{10.7937/gkr0-xv29}

\bibitem{psmapet2024dataset}
Gatidis, S., Kuestner, T.: A whole-body psma-pet/ct dataset with manually
  annotated tumor lesions (psma-pet-ct-lesions) (version 1) [dataset] (2024).
  \doi{10.7937/r7ep-3x37}

\bibitem{isensee2021nnu}
Isensee, F., Jaeger, P.F., Kohl, S.A., Petersen, J., Maier-Hein, K.H.: nnu-net:
  a self-configuring method for deep learning-based biomedical image
  segmentation. Nature methods  \textbf{18}(2),  203--211 (2021)

\end{thebibliography}

\end{document}